\documentclass[twocolumn]{aastex62}

 \usepackage[super]{nth}
\usepackage{color} 
\usepackage{bm}
\newcommand\modified[1]{{\bf #1}}

\newcommand\rband{$r\,$band}
\usepackage{graphicx}
\usepackage{natbib}

\accepted{ApJ, July 23, 2019}

\shorttitle{3-D Shape of XMPs}
\shortauthors{Putko et al.}

\begin{document}

\title{Inferring the 3-D shapes of extremely metal-poor galaxies from sets of projected shapes}

\author[0000-0002-1576-0083]{J.\,Putko}
\affiliation{Instituto de Astrof\'\i sica de Canarias\\
La Laguna, Tenerife, Spain}
\affiliation{Departamento de Astrof\'\i sica,\\
Universidad de La Laguna, Tenerife, Spain}

\author[0000-0003-1123-6003]{J.\,S\'anchez~Almeida}
\affiliation{Instituto de Astrof\'\i sica de Canarias\\
La Laguna, Tenerife, Spain}
\affiliation{Departamento de Astrof\'\i sica,\\
Universidad de La Laguna, Tenerife, Spain}

\author[0000-0001-8876-4563]{C.\,Mu\~noz-Tu\~n\'on}
\affiliation{Instituto de Astrof\'\i sica de Canarias\\
La Laguna, Tenerife, Spain}
\affiliation{Departamento de Astrof\'\i sica,\\
Universidad de La Laguna, Tenerife, Spain}

\author[0000-0002-1248-0553]{A.\,Asensio~Ramos}
\affiliation{Instituto de Astrof\'\i sica de Canarias\\
La Laguna, Tenerife, Spain}
\affiliation{Departamento de Astrof\'\i sica,\\
Universidad de La Laguna, Tenerife, Spain}

\author[0000-0002-1723-6330]{B.\,G.\,Elmegreen}
\affiliation{ IBM Research Division, T.J. Watson Research Center\\
1101 Kitchawan Road, Yorktown Heights\\
NY 10598, USA}

\author[0000-0002-1392-3520]{D.\,M.\,Elmegreen}
\affiliation{Department of Physics \& Astronomy, Vassar College\\
 Poughkeepsie, NY 12604, USA}


\begin{abstract}
The three-dimensional (3-D) shape of a galaxy inevitably is tied to how it has formed and evolved and to its dark matter halo. Local extremely metal-poor galaxies (XMPs; defined as having an average gas-phase metallicity $<$ 0.1 solar) are important objects for understanding galaxy evolution largely because they appear to be caught in the act of accreting gas from the cosmic web, and their 3-D shape may reflect this. Here we report on the 3-D shape of XMPs as inferred from their observed projected minor-to-major axial ratios using a hierarchical Bayesian inference model, which determines the likely shape and orientation of each galaxy while simultaneously inferring the average shape and dispersion.  We selected a sample of 149 XMPs and divided it into three sub-samples according to physical size and found that (1) the stellar component of XMPs of all sizes tends to be triaxial, with an intermediate axis $\approx\,$0.7 times the longest axis and that (2) smaller XMPs tend to be relatively thicker, with the shortest axis going from $\approx\,$0.15 times the longest axis for the large galaxies to $\approx\,$0.4 for the small galaxies.  We provide the inferred 3-D shape and inclination of the individual XMPs in electronic format. We show that our results for the intermediate axis are not clouded by a selection effect against face-on XMPs.  We discuss how an intermediate axis significantly smaller than the longest axis may be produced by several  mechanisms, including lopsided gas accretion, non-axisymmetric star formation, or coupling with an elongated dark matter halo. Large relative thickness may reflect slow rotation, stellar feedback, or recent gas accretion.

\end{abstract}

\keywords{
galaxies: dwarf --
galaxies: evolution --
galaxies: formation --
galaxies: fundamental parameters --
galaxies: irregular
}

\section{Introduction }

Simulations suggest that the main modes of galaxy growth over cosmic time involve merger \citep[e.g.,][]{2014ARA&A..52..291C} and cold gas streams (filaments of the cosmic web) that deeply penetrate the dark matter halo of galaxies and serve as the main driver of star formation \citep[e.g.,][]{2009Natur.457..451D}.  Although such cosmic accretion must become less intense as the Universe ages,  expands, and becomes less dense,  evidence of cosmic accretion has been mounting in the local Universe, as galaxies can be studied in greater detail nearby.  Local extremely metal-poor galaxies (XMPs; defined as having an average gas-phase metallicity $<$ 0.1 solar; e.g., \citeauthor{2000A&ARv..10....1K}~\citeyear{2000A&ARv..10....1K}) typically have several properties consistent with the cosmic accretion scenario: off-center clumps \citep[see Section~2.1; consistent with, e.g.,][]{1999ApJ...514...77N,2008ApJ...688...67E,2009ApJ...703..785D,2016MNRAS.457.2605C}; high specific star formation rates \citep[e.g.,][]{2011ApJ...743...77M}; isolated \citep{2015ApJ...802...82F}; rich in H\,I, especially in the outer regions \citep{2013A&A...558A..18F}; and significant metallicity drops at the starburst regions \citep{2013ApJ...767...74S,2015ApJ...810L..15S},
indicating recent accretion of metal-poor gas given that the timescale for gas mixing is short \citep[e.g.,][]{2002ApJ...581.1047D}.  Thus, in the context of cosmic accretion, XMPs are leading local laboratories for studying galaxy formation and evolution. 

About 500 XMPs have been identified thus far \citep[e.g.,][]{2017ApJ...835..159S}, and their three-dimensional (3-D) shape has yet to be explored.  The 3-D shape of a galaxy inevitably is linked to the past and/or current formation processes at work, so constraining the 3-D shape of XMPs may lead to a better understanding of the nature of primitive galaxies.  Specifically, it will provide insight regarding the roles of cosmic accretion and associated stellar feedback processes, the balance between external and internal processes, and the structural relation between the dark matter halo \citep[e.g.,][]{2015MNRAS.453..721V,2017MNRAS.467.3226V} and the stellar distribution, as dwarf galaxies, such as most XMPs, are likely dominated by dark matter \citep[e.g.,][]{2010ApJ...717..379B}. 

Here we infer the 3-D shape of XMPs using a simple and robust technique.  It is simple in the sense that it requires only fitting ellipses to galaxy images, and it is robust because the distribution of projected shapes is a consequence of the 3-D shape (and orientation) of the galaxies. Using an ellipsoid as our model for the 3-D shape, the 2-D projection in the plane of the sky is an ellipse. Thus, assuming the galaxies to have a common shape and to be oriented randomly, one can infer the parameters of the ellipsoid from the axial ratios of the ellipses fitted to their images.  We refer to this general technique as the $q$ \textit{technique}, where $q$ is the minor-to-major axial ratio of an ellipse fit to the image of a galaxy.   The $q$ technique has been used extensively starting with \citet{1926ApJ....64..321H}.  It is well suited for a hierarchical Bayesian inference model, and we use such an approach in this work (see Section~2.3).  Our inference method was inspired by \citet{2016ApJ...820...69S}, who emphasize the power of a Bayesian approach to the $q$ technique, namely that when using discrete $q$ measurements instead of a histogram of them, a large observed sample is not necessary and the uncertainty of the individual $q$ measurements can be used.  But our Bayesian approach is significantly different from theirs; ours is hierarchical, whereby the likely shape and orientation of each galaxy is inferred while simultaneously inferring the average shape and dispersion.

The paper is organized as follows: Section~2 describes our sample selection, $q$ measurements, and method for inferring 3-D shapes.  Section~3 presents our results and evaluates the potential bias caused by a surface brightness selection effect.  In Section~4, we discuss our main results.  Appendix~A highlights the link between the distribution of $q$ and 3-D shape through simulations, Appendix~B discusses the results of testing our inference model using a simulated galaxy population and small samples, and Appendix~\ref{app:selection} shows how to modify the simple 3-D model used for shape to account for the surface brightness variation with $q$.

\section{Methods}

\subsection{Sample Selection}\label{sec:sample_selection}

\begin{figure}
\includegraphics[scale = 0.6]{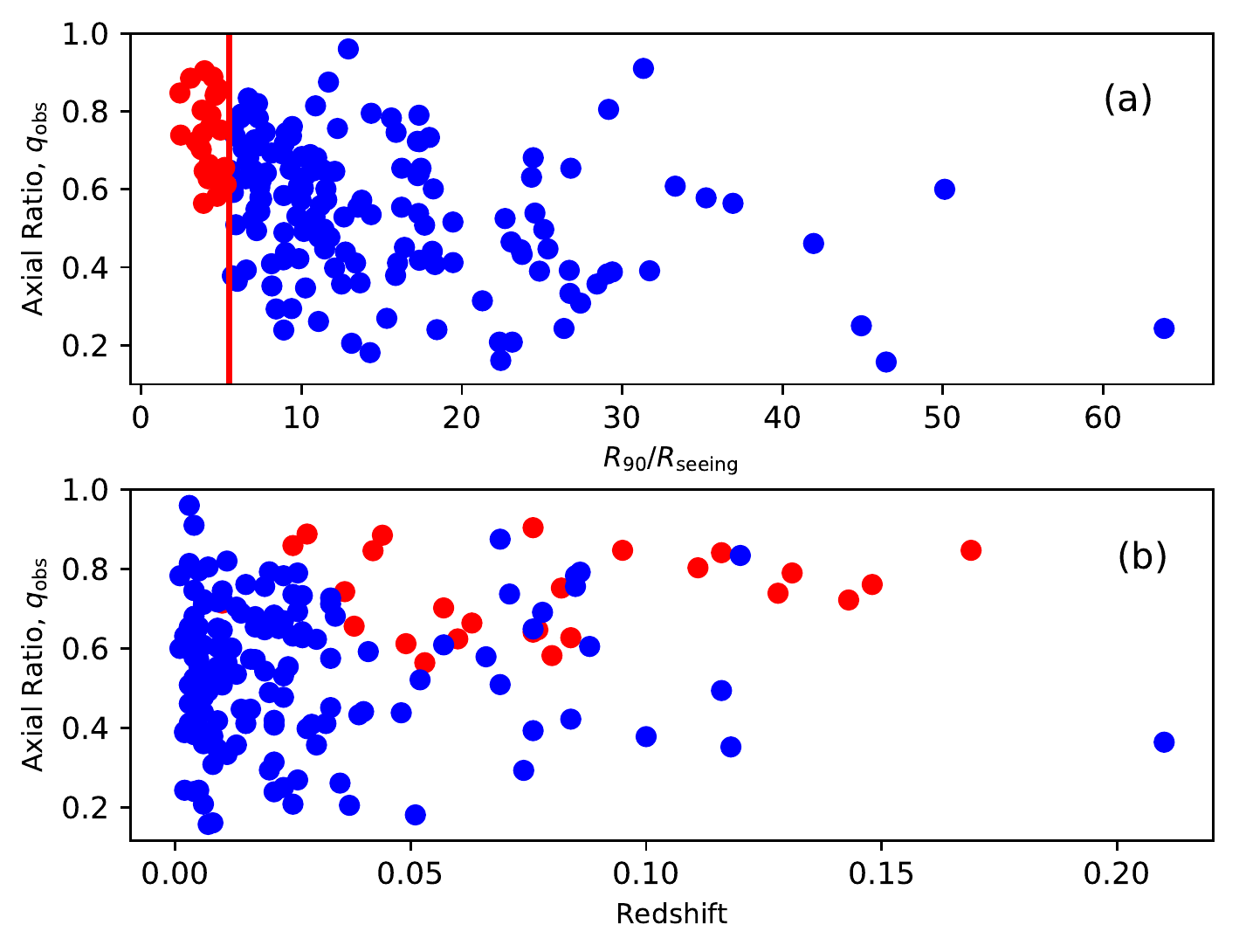} 
\caption{
(a) Observed axial ratio versus  galaxy angular size (radius enclosing 90\% of the light in the \rband) relative to seeing (HWHM of the PSF) for the XMPs. In order to minimize the influence of seeing, XMPs with $R_{90} < 5 R_{\rm seeing}$ are discarded from the analysis (those to the left of the vertical line). 
(b) Observed axial ratio versus redshift, showing no obvious trend. The color code is the same as in panel (a), where red corresponds to $R_{90} < 5 R_{\rm seeing}$.
}
\label{fig:new_fig1}
\end{figure}

The galaxies used in our inference of 3-D shape are from \citet{2016ApJ...819..110S}, who mined the spectroscopic catalog of the Sloan Digital Sky Survey (SDSS) in producing the largest published sample (195) of XMPs from a single survey.  To have a more homogeneous sample, we omitted 4 spiral galaxies (showing spiral arms and/or a bulge), 6 interacting galaxies with obvious tidal tails or bridges between separate nuclei, and 5 galaxies with a non-elliptical projected shape. 
The size of the remaining 180 galaxies relative to seeing is represented in Fig.~\ref{fig:new_fig1}, top panel. It shows the observed axial ratio (computed in Sect.~\ref{sec:qmeasurement}) versus $R_{90} / R_{\rm seeing}$. $R_{90}$ is the equivalent radius enclosing 90\% of the light ($\pi R_{90}^2$ is the area enclosing 90\% of the light) whereas $R_{\rm seeing}$ stands for the half width at half maximum (HWHM) of the seeing point spread function (PSF). Both $R_{90}$ and $R_{\rm seeing}$ were taken from the SDSS database and correspond to the \rband.  XMPs with $R_{90} \sim\ R_{\rm seeing}$ tend to be round (i.e., with axial ratio close to one), which can be an artifact produced by seeing.  The trend disappears at $R_{90} > 5 \, R_{\rm seeing}$ (Fig.~\ref{fig:new_fig1}, top panel), therefore, we further omitted the 31 galaxies with $R_{90} < 5 \, R_{\rm seeing}$; this discarded compact objects appearing as clumps without a host, along with other poorly resolved galaxies.  The final selected sample contains 149 XMPs. We note that the cut in size may potentially introduce an additional bias against physically small rounded galaxies (i.e., those to the left of the vertical line in Fig.~\ref{sec:qmeasurement}, top panel), however, the analysis carried out in Sect.~\ref{sec:shape} proves that it has insignificant impact on the inferred galaxy shapes.  

%

Among our selected sample, the majority have at least one clump appearing as a distinct object within a host (the clumps are off-center in all but 2 cases), a large fraction have an off-center bluer and/or brighter region (not appearing as a distinct object), and a small fraction show no signs of \textit{clumpiness} but still appear to lack structure.  We used the SDSS SkyServer\footnote{http://skyserver.sdss.org} RGB images for this visual assessment of clumpiness, but clumpiness was not part of our selection criteria.  Galaxy redshifts go from 0 to 0.2, with most XMPs being closer than 0.05,  and with no obvious trend between axial ratio and redshift \modified{once the galaxies with  $R_{90} < 5 R_{\rm seeing}$ are discarded (see the blue symbols in } Fig.~\ref{fig:new_fig1}, bottom panel).

\subsection{Axial Ratio Measurement}\label{sec:qmeasurement}

We used the \rband\ SDSS images of data release 12 \citep{2015ApJS..219...12A} for our observed $q$ ($q_{{\rm obs}}$) measurements. The \rband\ is the deepest and traces old stellar populations that we expect most XMPs to contain \citep[e.g.,][]{2008ApJ...675..194C}.  We used the IRAF\footnote{IRAF (Image Reduction and Analysis Facility) is distributed by the National Optical Astronomy Observatories, which are operated by the Association of Universities for Research in Astronomy, Inc., under cooperative agreement with the National Science Foundation.} task \textit{ellipse} \citep{1996ASPC..101..139B} to derive our $q_{{\rm obs}}$ measurements from elliptical isophotes.  Our ellipsoid model for 3-D shape (see Section~2.3) implies that its 2-D projection has an elliptical shape given by a single center ($X$, $Y$), position angle ($PA$), and $q$ value, so we ran \textit{ellipse} twice: first allowing $X$, $Y$, $PA$, and $q$ to vary among each galaxy's isophotes and then with $X$, $Y$, and $PA$ fixed to the median resulting from the first run and allowing only $q$ to vary.  In both runs, we used only the isophotes in the range $22.5-24.5$ mag arcsec$^{-2}$ because this samples primarily the outskirts of the galaxies, where (1) $q$ is less affected by seeing, (2) clumps are more likely to be avoided, (3) $q$ is less uncertain in the sense that more pixels are used in the fitting, and (4) $q$ tends to vary less between successive isophotes.  Also in both runs, we clipped strongly deviant pixels (e.g., clumps), we used small spacing between successive isophotes to generate a large number of $q$ measurements, and we discarded isophotes whose fits did not converge.

According to our ellipsoid model, the different $q$ values given by the \nth{2} set of isophotes for each galaxy are all equally valid, so we took the median to give $q_{{\rm obs}}$, and we took the standard deviation to give $\sigma_{q_{{\rm obs}}}$, the uncertainty in $q_{{\rm obs}}$.  Our $\sigma_{q_{{\rm obs}}}$ estimates are in the interval [0.01, 0.11], and the median is 0.04. The effect of seeing is not included in $\sigma_{q_{{\rm obs}}}$, but its impact should be small given our cut in galaxy size. We have estimated that even for the 25\,\% smallest galaxies it would be only 0.03.

We checked the \textit{ellipse} fits by eye against smoothed contour plots.  The starting semi-major axis and the clipping parameters were tweaked by trial-and-error if a fit failed due to a clump or foreground star interfering.

\subsection{Inferring 3-D Shape}\label{sec:inferring_shape}

\begin{figure}
\plotone{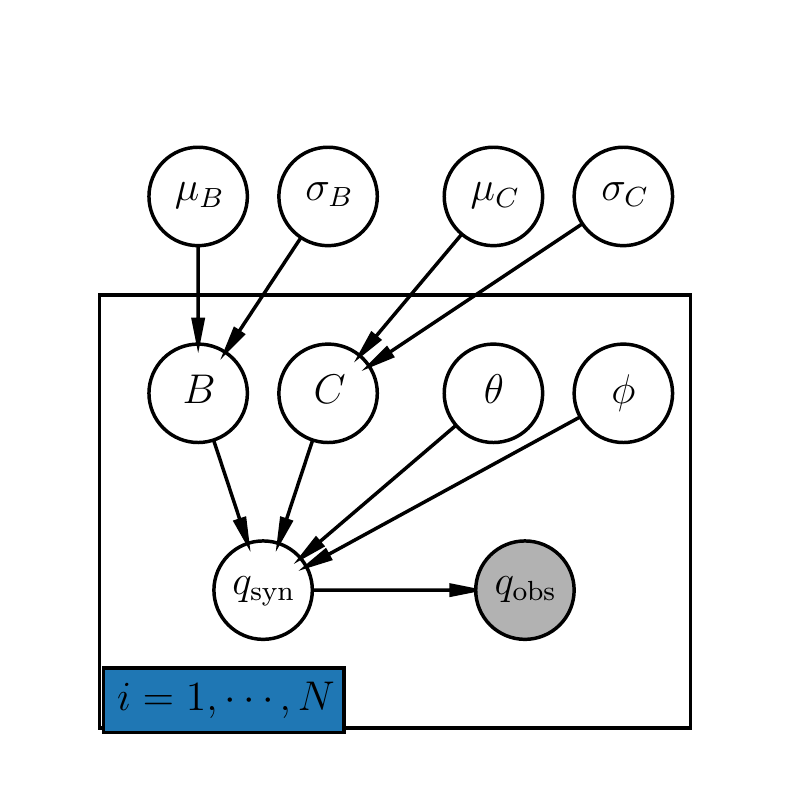}
\caption{A graphical model summarizing the conditional dependencies among the parameters in our inference model. The box is repeated for $N$ observed galaxies, with all boxes connected to $\mu_{B}$, $\sigma_{B}$, $\mu_{C}$, and $\sigma_{C}$.  See Section~2.3 for the definition of each parameter. \label{fig-daft}}
\end{figure}

Here we explain our hierarchical Bayesian inference model to carry out the $q$ technique; that is, to infer the 3-D shape distribution of our sample from our $q_{{\rm obs}}$ measurements by modeling the galaxies as ellipsoids oriented randomly.  Our inference model intertwines two stages, and Figure~\ref{fig-daft} shows the conditional dependencies involved (described below).  The shape of each galaxy is inferred (stage 1) while inferring the global parameters characterizing the distribution of shapes (stage 2). The Bayesian solution to our statistical problem is known as the \textit{posterior} probability\footnote{Probability of the model parameters given the observations.}, and it is proportional to some \textit{likelihood function}\footnote{Probability of the observations given the model parameters.} times some \textit{prior} probability.\footnote{Probability of the model parameters before considering the observations.}  

To compute the posterior, one needs to define the likelihood function, which requires a \textit{generative model} for modeling the probability of $q_{{\rm obs}}$ given $q_{{\rm syn}}$, a synthetic $q$ value.  We assume the observed galaxies to have an ellipsoidal form with the mutually perpendicular axes length ($A$), width ($B$), and thickness ($C$) normalized to $A$, therefore leading to a ratio of axes 1:$B$:$C$, where $A=1 \geq B \geq C > 0$.  Thus, any ellipsoid shape is defined by $B$ and $C$ alone.  The ellipsoid orientation depends on the inclination angle $\theta$ and the azimuthal angle $\phi$.  We used \citet{1998NCimB.113..927S} to compute $q_{{\rm syn}}$ for any \{$B$, $C$, $\theta$, $\phi$\} set of values; for the projected semi-major ($a$) and semi-minor ($b$) axes, they provide the relations
\begin{equation}
a^{2} b^{2}=f^2=
\label{eq:area}
\end{equation}
\begin{displaymath}
\left(C\sin\theta\cos\phi\right)^{2}+\left(BC\sin\theta\sin\phi\right)^{2}+\left(B\cos\theta\right)^{2}
\end{displaymath}
and
\begin{equation}
a^2+b^2 = g = \cos^2\phi+\cos^2\theta\sin^2\phi + 
\end{equation}
\begin{displaymath}
B^2(\sin^2\phi+\cos^2\theta\cos^2\phi)+C^2\sin^2\theta,
\end{displaymath}
so that, defining $h$ as 
\begin{equation}
h=\sqrt{\frac{g-2f}{g+2f}}\;,
\end{equation}
we can express $q_{{\rm syn}}$ as
\begin{equation}
q_{{\rm syn}}=\frac{b}{a}=\frac{1-h}{1+h}\;.
\label{eq:qsyn}
\end{equation}
To highlight the link between $q$ and 3-D shape, Figure~\ref{fig-demo} in Appendix~A shows  histograms of $q_{{\rm syn}}$ for randomly oriented galaxies and a variety of 3-D shapes.

Under the assumption of uncorrelated Gaussian noise, our generative model for sampling the posterior is $q_{{\rm syn}}(B,C,\theta,\phi)+\epsilon=q_{{\rm obs}}$, where the noise contribution $\epsilon$ is a Gaussian-distributed random variable, with mean zero and standard deviation $\sigma_{q_{{\rm obs}}}$, that is truncated and renormalized because $q$ must be in the interval (0,1].  We will refer to \{$B$, $C$, $\theta$, $\phi$\} as the set of \textit{galaxy parameters} whose priors are further governed by the assumption that the observed galaxies make up a family of ellipsoids defined by the \textit{family parameters} $\mu_{B}$, $\mu_{C}$,  $\sigma_{B}$, and $\sigma_{C}$. Thus the galaxy parameters of a set of galaxies, $\vec{B}$ and $\vec{C}$\footnote{Vector notation is used to represent a set of galaxy parameters for a set of galaxies.}, are assumed to follow from the same Gaussian distributions with means $\mu_{B}$ and $\mu_{C}$, and standard deviations $\sigma_{B}$ and $\sigma_{C}$.


The prior of our statistical problem is the product of the priors indicated in Table 1.  We now have all the ingredients to write the expression for evaluating the posterior (Bayes' theorem) in terms of the parameters in our inference model:
\begin{equation}
P({\vec{B}},{\vec{C}},\vec{\theta},\vec{\phi},\mu_{B},\sigma_{B},\mu_{C},\sigma_{C}\mid\vec{q_{{\rm obs}}})
\propto
\end{equation}
\begin{displaymath}
\mathcal{L}(\vec{q_{{\rm obs}}}\mid{\vec{B}},{\vec{C}},\vec{\theta},\vec{\phi})\, P({\vec{B}}\mid\mu_{B},\sigma_{B}) \,P({\vec{C}}\mid\mu_{C},\sigma_{C})
\end{displaymath}
\begin{displaymath}
P(\vec{\theta})\,P(\vec{\phi})\,P(\mu_{B})\,P(\mu_{C})\,P(\sigma_{B})\,P(\sigma_{C})\;, 
\end{displaymath}
where $P(y|x)$ stands for the probability of $y$ given $x$ and $\mathcal{L}$ is substituted for $P$ to denote the likelihood.  Note that the family parameters appear in the likelihood function implicitly.    
\begin{deluxetable}{lll}
\tablecolumns{3}
\tablewidth{0pc}
\tablecaption{Priors\label{tbl-priors}}
\tablehead{
		\colhead{Parameter}   & 
		\colhead{Probability Distribution\tablenotemark{$b$}}   & 
		\colhead{Type}
		}  
\startdata
cos\,$\theta$ & $U($-$1, 1)$ & prior \\
$\phi$ & $U(0, 2\pi)$ & prior \\
$B\tablenotemark{$a$}$ & $N(\mu_{B}, \sigma_{B})$ & hierarchical prior \\
$C\tablenotemark{$a$}$ & $N(\mu_{C}, \sigma_{C})$ & hierarchical prior \\
$\mu_{B}$ & $U(0, 1)$ & hyperprior \\
$\mu_{C}$ & $U(0, 1)$ & hyperprior \\ 
$\sigma_{B}$ & \textit{Half}-$N(0.05)$ & hyperprior \\
$\sigma_{C}$ & \textit{Half}-$N(0.05)$ & hyperprior \\
\enddata
\tablenotetext{a}{$B$ and $C$ have the additional constraints $1 \geq B \geq C > 0$.}
\tablenotetext{b}{$U$ stands for uniform. $N$ stands for normal.  Given a normally-distributed random variable $X$, the random variable $Y = |X|$ has a half-normal distribution \textit{Half}-$N$.}
\end{deluxetable}

Our method to evaluate the posterior is computationally efficient for three reasons, and the first two are consequences of our hierarchical framework.  First, we have circumvented computing $q_{{\rm syn}}$ histograms corresponding to different \{$\mu_{B}$, $\sigma_{B}$, $\mu_{C}$, $\sigma_{C}$\} sets.  Second, only parameter estimates near the maximum likelihood will be sampled.  And third, we implemented our inference model using Stan\footnote{Stan is written in C++  (see mc-stan.org).}, which uses  the state-of-the-art sampling algorithms NUTS \citep{HoffmanandGelman2014}. 
 We used Stan also because it makes relatively straightforward to account for the change of variables needed to ensure the condition $B \geq C$.

We checked our inference model in three different ways:
(1) Although the NUTS algorithm implemented in Stan is known to work properly even in problematic hierarchical inference models, we checked that the sampling of the posterior properly converged by checking standard convergence criteria\footnote{Such as the number of effective samples and the potential scale reduction statistics (see Stan manual).}. (2) \emph{Posterior predictive checks}  (explained in Section 3.1) were carried out to be sure that our inference model is able to properly explain the observations. (3) We also fed model populations to our inference model to see how well it can infer the family parameters and how sample size plays a role. We found that indeed $\mu_{B}$ and $\mu_{C}$ can be accurately inferred from small samples and that $\sigma_{B}$ can be particularly difficult to constrain.  Appendix~B presents the experiment and results in more detail, and Figure~\ref{fig-sigma} is another way to argue that $\sigma_{B}$ can be difficult to constrain.

\section{Results}

\begin{figure*}
\plotone{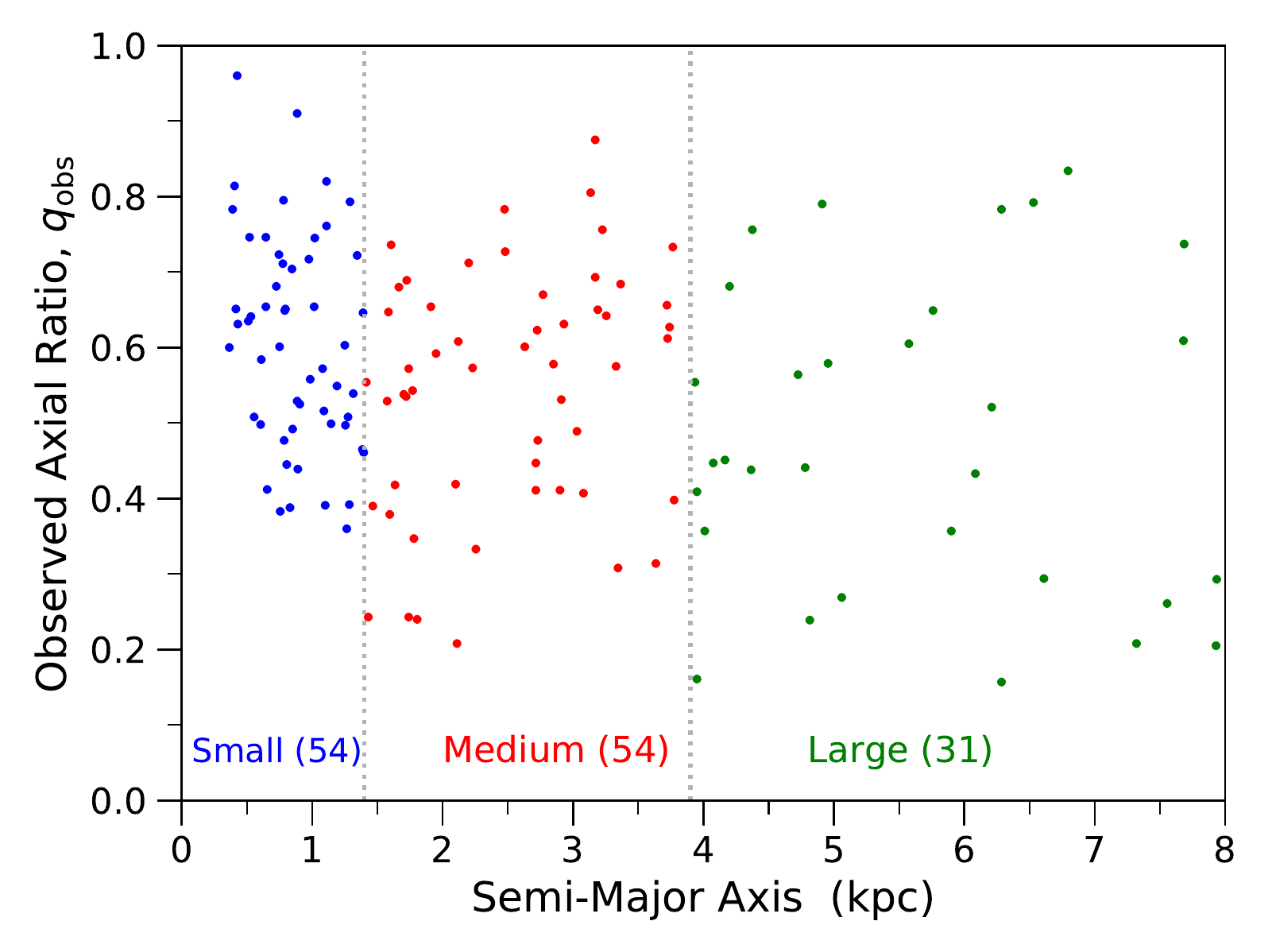}
\caption{Our $q_{{\rm obs}}$ values versus semi-major axis in the \rband.  We divided the full sample into three sub-samples (number of objects in parentheses), indicated by the gray dashed lines at 1.4 and 3.9 kpc, and we chose to omit galaxies larger than 8 kpc from the sample Large.  The color scheme matches with Figures~\ref{fig-clouds} and~\ref{fig-sb}.\label{fig-cuts}}
\end{figure*}

Relative thickness $C$ is directly signaled by the lowest $q$ values, so we can already infer from Figure~\ref{fig-cuts}, showing $q_{{\rm obs}}$ versus semi-major axis, that smaller galaxies tend to be relatively thicker (larger $C$).  We therefore divided our sample into {\em Small}, {\em Medium}, and {\em Large}, as shown in Figure~3.  (To convert to physical size, we estimated distance using the SDSS redshift of each galaxy and a Hubble constant of 70 km\,s$^{-1}$\,Mpc$^{-1}$.  Only 4\% of our sample is nearer than 10 Mpc --- shortly beyond this point, proper motions with respect to the Hubble flow become negligible --- so the vast majority of our size measurements are reasonable approximations.)  We made the 3 cuts marked by the gray dashed lines in Figure~\ref{fig-cuts}.  The cut positions were chosen at semi-major axis of 1.4~kpc (so that $q_{{\rm obs}} > 0.3$ for all Small; see Figure~3), 3.9~kpc (so that $q_{{\rm obs}} > 0.2$ for all Medium), and 8~kpc.  
Ten galaxies larger than 8 kpc (up to 23 kpc) were not included in Large to have a more homogeneous sub-sample, though we noted that this does not impact the results significantly.
In addition to these three cuts, we also use the whole sample of 149 galaxies, naming it {\em All}. 

\subsection{Observed $q$ Histograms and Posterior Predictive Checks}

\begin{figure*}
\plotone{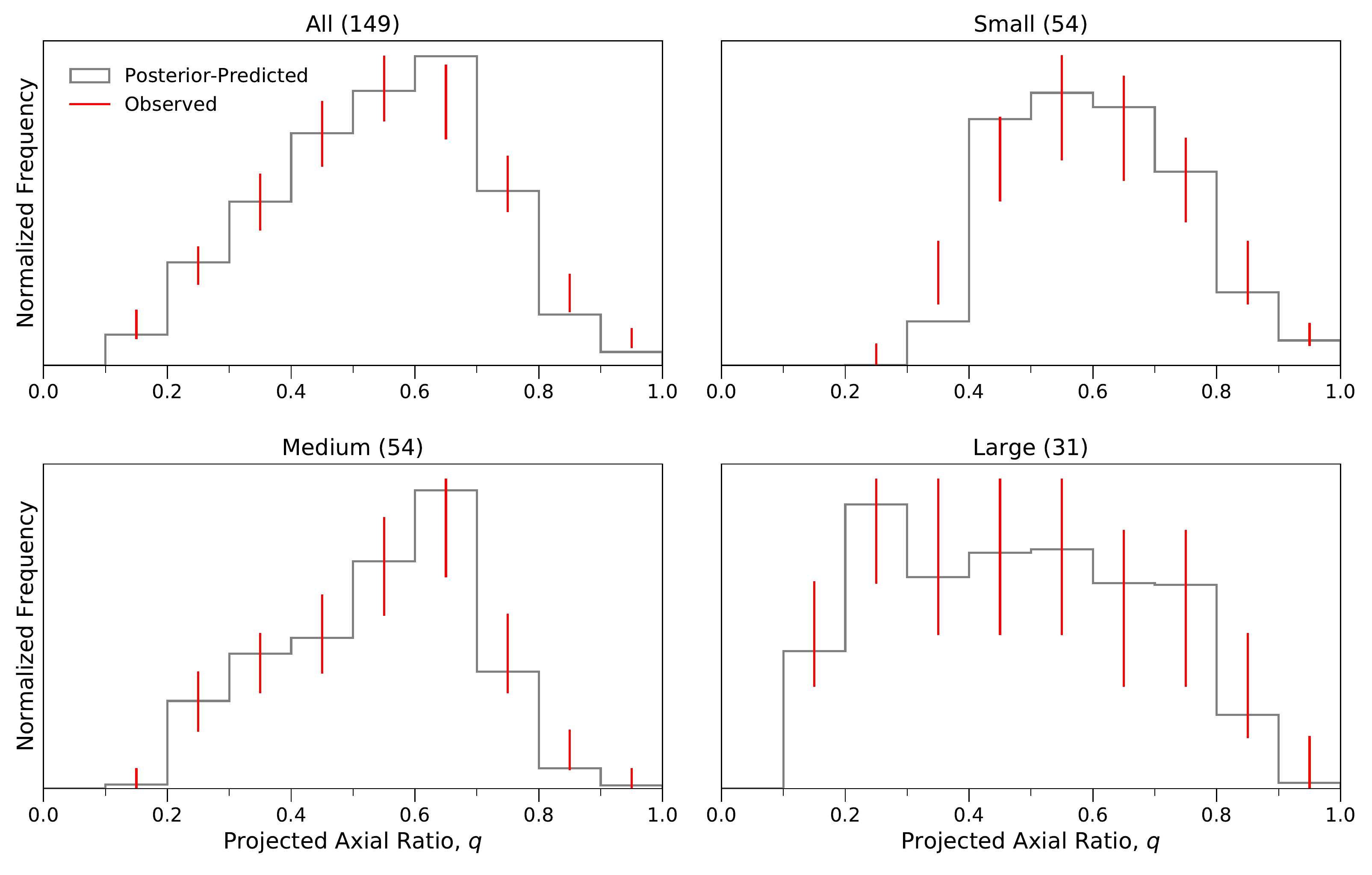}
\caption{In red: Our $q_{{\rm obs}}$ histograms and their 1$\sigma$ uncertainty estimated via Monte Carlo simulation. The bin size is 0.1.   In gray: The posterior-predicted $q_{{\rm syn}}$ histograms (see Section~2.3).  The name of each sample is shown atop each panel with sample size in parentheses. \label{fig-ppc}}
\end{figure*}

The inferred 3-D shapes to be presented in Section~3.2 are based on the discrete $q_{{\rm obs}}$ values and their uncertainties, but first we show our $q_{{\rm obs}}$ histograms in Figure~\ref{fig-ppc} and their posterior-predicted fits to check that our inference model worked as expected.  Our $q_{{\rm obs}}$ histograms and their 1$\sigma$ uncertainty are represented by the red lines, generated via Monte Carlo simulation.  The posterior-predicted histograms in gray show the $q_{{\rm syn}}$ values corresponding to the marginal posterior sampling of the galaxy parameters (see Section~2.3).  The agreement between our observed distributions and the posterior-predicted histogram shows that our inference model is appropriate for explaining  the observations.

\subsection{Inferred 3-D Shapes}\label{sec:shape}

\begin{figure*}
\plotone{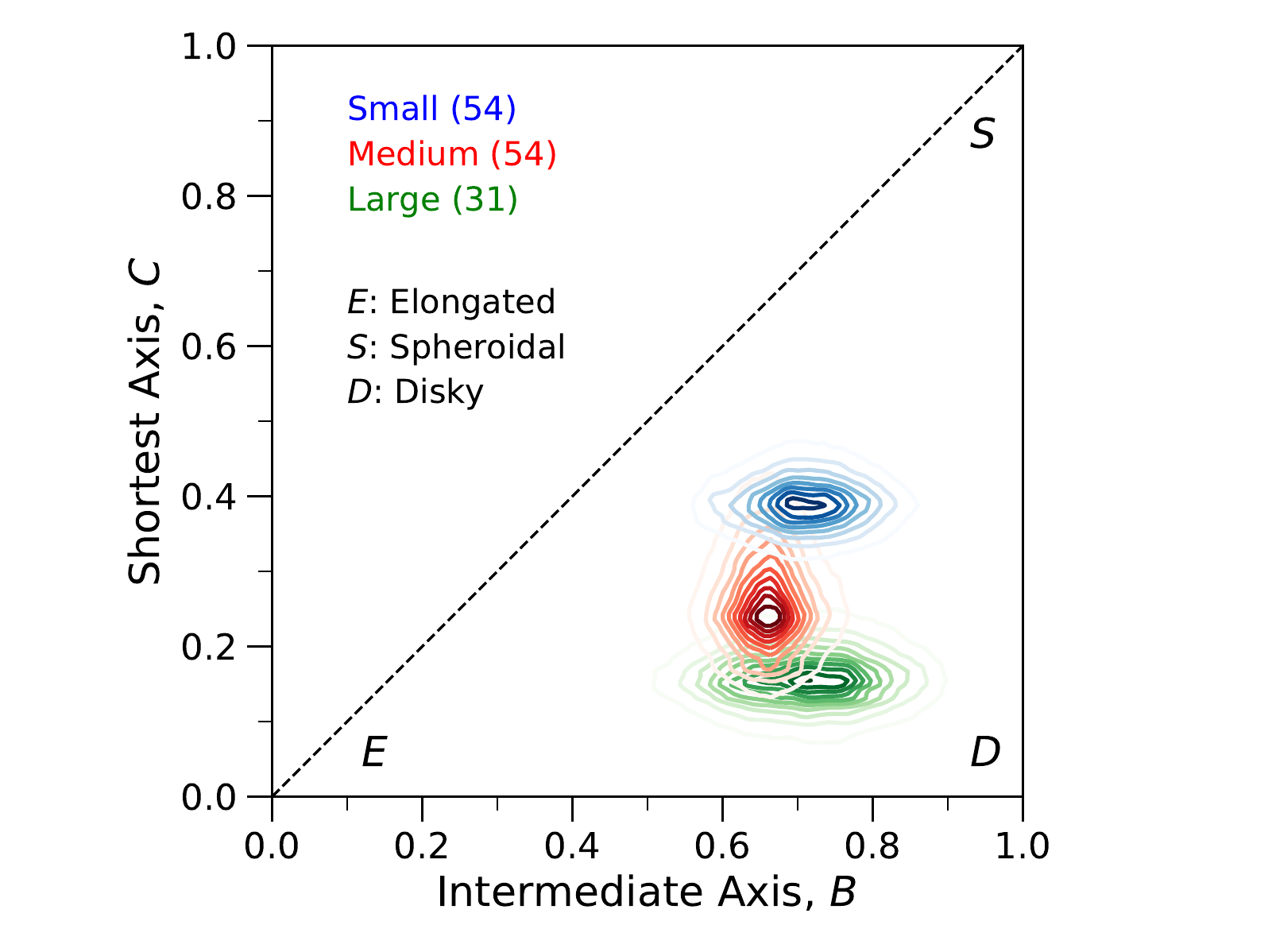}
\caption{The inferred shape distributions of our sub-samples (number of objects in parentheses), based on the marginal posteriors of $C$ and $B$ for each galaxy.  Darker represents higher probability according to a linear scale.  The color scheme matches with Figures~\ref{fig-cuts} and~\ref{fig-sb}. The longest ellipsoid axis is always 1, and all possible ellipsoidal shapes satisfy the condition $1 \geq B \geq C > 0$ (i.e., no shapes exist above the dashed line).  Closer to corner $E$ is more elongated, closer to corner $S$ is more spheroidal, and closer to corner $D$ is more disky.\label{fig-clouds}}
\end{figure*}

The inferred 3-D shape distributions for our 3 sub-samples are shown in Figure~\ref{fig-clouds}.  We plot the  marginal posterior sampling of $\vec {C}$ versus that of $\vec{B}$ and translate the density of points to a color map where darker represents higher probability.  In other words, each galaxy has its own \textit{cloud} in the $C$ (relative thickness) versus $B$ (relative width) parameter space that is its inferred shape and the uncertainty in it, and the clouds in Figure~\ref{fig-clouds} combine all of the individual galaxy clouds into a family cloud.  The maximum a posteriori value of each family parameter is given in Table 2 for each of our sub-samples.   The shape parameters of the individual galaxies are included in Table~\ref{tbl-galparams}, which is given on-line.  The shape distributions in Figure~\ref{fig-clouds} offer a more complete view because the $\vec{C}$ and $\vec{B}$ marginal posteriors inevitably are not perfectly Gaussian.  The $\sigma_{B}$ and $\sigma_{C}$ results in Table 2 are poorly constrained (this is inherent to the inference model; see Appendix~B), but we see that the average 3-D shape of each sub-sample is nevertheless well-constrained.

\begin{deluxetable*}{lccccc}
\tablecolumns{8}
\tablewidth{0pc}
\tablecaption{Maximum A Posteriori Family Parameters\tablenotemark{$a$}\label{tbl-results}}
\tablehead{
		\colhead{Sample\tablenotemark{$b$}}   & 
		\colhead{Sample Size}   & 
		\colhead{$\mu_{\rm C}$}    &
		\colhead{$\sigma_{\rm C}$}    &
		\colhead{$\mu_{\rm B}$}    &
		\colhead{$\sigma_{\rm B}$} }  
\startdata
Small & 54 & $0.39^{+0.02}_{-0.02}$ & $0.03^{+0.02}_{-0.02}$ & $0.71^{+0.05}_{-0.05}$ & $0.04^{+0.03}_{-0.02}$ \\
Medium & 54 & $0.26^{+0.04}_{-0.04}$ & $0.06^{+0.03}_{-0.04}$ & $0.66^{+0.03}_{-0.03}$ & $0.03^{+0.02}_{-0.02}$ \\
Large & 31 & $0.16^{+0.03}_{-0.03}$ & $0.03^{+0.02}_{-0.02}$ & $0.71^{+0.08}_{-0.08}$ & $0.04^{+0.03}_{-0.02}$ \\
All & 149 & $0.28^{+0.02}_{-0.02}$ & $0.09^{+0.02}_{-0.02}$ & $0.67^{+0.03}_{-0.03}$ & $0.03^{+0.02}_{-0.02}$ \\
\enddata
\tablenotetext{a} {We report the median values of the marginal posteriors and express the uncertainty using the 16th and 84th percentiles (the probability that the parameter is within this range is 68\%).}
\tablenotetext{b}{The sub-samples are divided as shown in Figure 3. The sub-sample All also includes the 10 galaxies excluded in Large.}
\end{deluxetable*}
\begin{deluxetable*}{ccccccc}
\tablecolumns{9}
\tablewidth{0pc}
\tablecaption{3-D shape parameters for individual galaxies\tablenotemark{$a$}\label{tbl-galparams}}
\tablehead{
		\colhead{R.A.\tablenotemark{$b$}}   & 
		\colhead{Dec.\tablenotemark{$b$}}   & 
		\colhead{$C$}    &
		\colhead{$B$}    &
		\colhead{$\theta$\tablenotemark{$c$}}    &
		\colhead{Sub-Sample\tablenotemark{$d$}}    &
		\colhead{ID\tablenotemark{$e$}} \\
                \colhead{(deg)}&\colhead{(deg)}&\colhead{(A)}&\colhead{(A)}&(deg)& --&--
}
\startdata
9.42130 & 0.55561 & $0.39^{+0.04}_{-0.04}$ & $0.72^{+0.07}_{-0.07}$ & $43.0^{+20.8}_{-20.1}$ & S & 6  \\
19.80951 & -9.59617 & $0.37^{+0.03}_{-0.04}$ & $0.71^{+0.07}_{-0.07}$ & $81.8^{+5.6}_{-8.5}$ & S & 9  \\
23.46897 & 13.70264 & $0.39^{+0.04}_{-0.04}$ & $0.71^{+0.07}_{-0.07}$ & $49.2^{+22.1}_{-22.5}$ & S & 10  \\
45.45427 & -0.88260 & $0.39^{+0.04}_{-0.04}$ & $0.71^{+0.07}_{-0.07}$ & $72.5^{+12.0}_{-13.9}$ & S & 19  \\
130.65241 & 10.55387 & $0.39^{+0.04}_{-0.04}$ & $0.71^{+0.07}_{-0.08}$ & $69.6^{+13.8}_{-13.3}$ & S & 36  \\
\vdots & \vdots & \vdots & \vdots &\vdots & \vdots & \vdots \\ 
\enddata
\tablenotetext{a} {The full table, available only online, contains 139 galaxies divided among our three sub-samples.  It is sorted by sub-sample and then by R.A.  The corresponding family parameters are described in Section 2.3 and Table~\ref{tbl-results}.  We report the median values of the marginal posteriors and express the uncertainty using the 16th and 84th percentiles (the probability that the parameter is within this range is 68\%).}
\tablenotetext{b}{R.A.\ and Dec.\ in J2000 coordinates.} 
\tablenotetext{c}{Inclination of the major axis with respect to the line-of-sight.}
\tablenotetext{d}{Figure~\ref{fig-cuts} shows how sub-samples are divided. S, M and L stand for Small, Medium, and Large, respectively.}
\tablenotetext{e}{Identification number from S\'anchez Almeida et al.\ (2016).}
\end{deluxetable*}

In Figure~\ref{fig-clouds} we see the relation that relative thickness increases with decreasing galaxy size.  We see no trend in the results for relative width, but we see that for all three sub-samples, the galaxies are inferred to be notably far from axisymmetric disks.  In Figure~\ref{fig-clouds} we have included a variation of the intuitive-shape framework (elongated vs.\ disky vs.\ spheroidal) proposed by \citet{2014ApJ...792L...6V}.

In order to avoid the effect of seeing on the inferred shapes, we removed from the original sample 31 galaxies with size $R_{90} < 5 \, R_{\rm seeing}$ (Sect.~\ref{sec:sample_selection}). This cut in size may potentially introduce a bias against physically small round galaxies. In order to discard it, we repeated the computation of 3-D shape including all the small filtered out galaxies for which $q$ could be estimated.  The new values for $B$ and $C$ are very close to those in Table~\ref{tbl-results}, namely, $\mu_C$ and $\mu_B$ are 
$0.39^{+0.02}_{-0.02}$ and $0.73^{+0.05}_{-0.07}$,
$0.29^{+0.04}_{-0.04}$ and $0.67^{+0.04}_{-0.04}$,
and $0.17^{+0.03}_{-0.03}$ and $0.79^{+0.05}_{-0.05}$,
for the sub-samples small, medium, and large, respectively.

\subsection{Surface Brightness Selection Effect}\label{sec:selection}

Disk-like galaxies tend to have fainter surface brightness when observed face-on.  A drawback of our implementation of the $q$ technique is that it does not account for the bias introduced by this effect, i.e., for the fact that near the surface brightness limit of a survey different orientations have different probabilities of being observed. To assess the potential impact of this effect on our results, we estimate the expected surface brightness bias. As we will show, the predicted decrease in surface brightness when $q$ increases is not present in the observed data set. Thus, the surface brightness bias does not seem to be important for the XMP galaxies analyzed in our work, supporting  the reliability of the shapes inferred using our implementation of the $q$ technique.

\begin{figure*}
\plotone{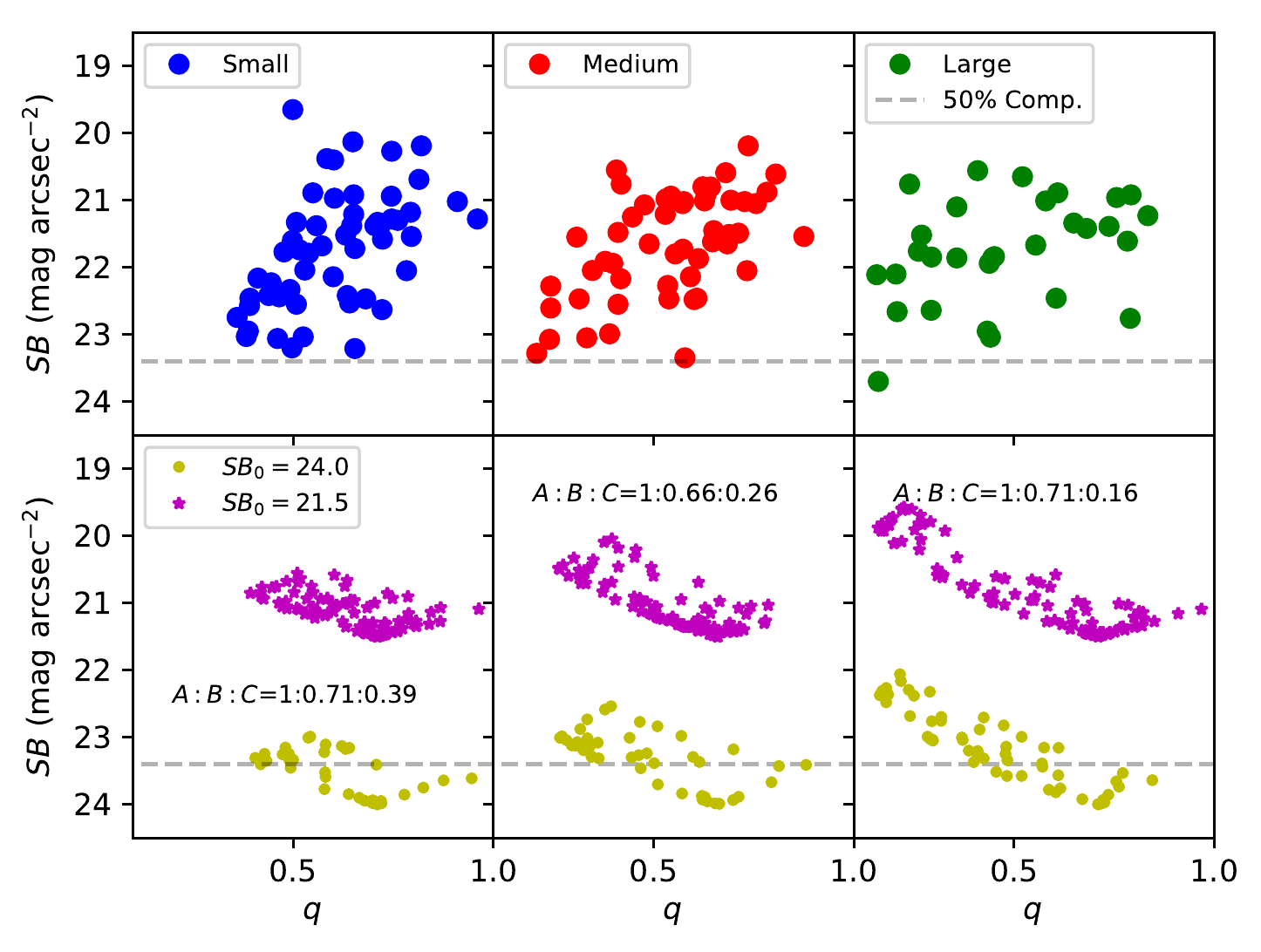}
\caption{Top row: half-light surface brightness in the \rband\ versus $q$ for our sub-samples Small, Medium, and Large, as labelled. The dashed lines show the 50\,\%  completeness level for the SDSS spectroscopic catalog, as reported by \citet{2005ApJ...631..208B}.  The color scheme matches with Figures~\ref{fig-cuts} and~\ref{fig-clouds}. Bottom row: a model for the bias due to the surface brightness selection effect considering different face-on brightnesses and galaxy shapes (see Section~3.3 for further details).  Each point corresponds to a galaxy oriented randomly,  with the axial ratio $A:B:C$ and lowest surface brightness $SB_0$ given in the insets. \label{fig-sb}}
\end{figure*}

In this estimate, we crudely model the surface brightness of a galaxy ($SB$) assuming no internal extinction. (Extinction is known to be low in XMPs, and will be treated in Appendix~\ref{app:selection}.) Then the total flux emitted by the galaxy is independent of its orientation; therefore,  the flux per unit surface scales as the inverse of the area projected by the galaxy on the plane of the sky. Assuming the 3-D model used in the paper, this area is proportional to the product $ab$ given in Equation~(\ref{eq:area}), so that
\begin{equation}
SB = SB_0 -2.5\,\log\big[B/(ab)\big],
\label{eq:myeq}
\end{equation}  
where $SB_0$ stands for the largest $SB$ considering all posible orientations of the galaxy, and $B$ is the maximum value of $a b$\footnote{The fact that $a b \leq B$ follows from Equation~(\ref{eq:area}). The partial derivates of $a b$ with respect to $\theta$ and $\phi$ are zero at the maximum, which is achieved only when $\theta=0$, and so,  when $ab = B$.}.
Given $B, C, \theta, \phi$ and $SB_0$, $SB(q)$ follows implicitly from Equations (\ref{eq:area}), (\ref{eq:qsyn}), and (\ref{eq:myeq}). Figure~\ref{fig-sb}, bottom row, shows scatter plots of $SB$ versus $q$ for collections of 100 identical galaxies randomly oriented. (Their actual $B, C$ and $SB_0$ values are given in the figure.) In order to account for the limited completeness of our survey, the 100 model galaxies were included in Figure~\ref{fig-sb} at random, with the probability depending on their $SB$. The probability of being included was given by the completeness function of the SDSS spectroscopic survey worked out by \citet{2005ApJ...631..208B}, who added mock disk galaxies to the raw SDSS images, which then went through the standard pipeline to be selected as member of the SDSS spectroscopic catalog. The completeness function thus estimated has 50\% completeness at 23.4\,mag\,arcsec$^{-2}$ and fall-off width of 
around 1\,mag\,arcsec$^{-2}$. Thus, the selection effect is not noticeable until $SB_0=24$\,mag\,arcsec$^{-2}$, but it is severe for $SB_0=25$\,mag\,arcsec$^{-2}$ and larger. 

The model distribution $SB(q)$ in Figure~\ref{fig-sb} (bottom row) have been chosen to have $B$ and $C$ as the means $\mu_B$ and $\mu_C$ observed in the samples Small, Medium and Large (Table~\ref{tbl-results}). They have to be compared with the observed distribution in Figure~\ref{fig-sb}, top row. For our observed $SB$ distributions in Figure~\ref{fig-sb} we plot the half-light surface brightness in the \rband, as this is the quantity  employed by \citet{2005ApJ...631..208B} to quantify the SDSS completeness function used in our modeling. Comparing our model with observed half-light surface brightness implicitly assumes the galaxies to have a single shape, from the isophotes traced by the half-light radius to those where $q$ is measured. We used SExtractor \citep{1996A&AS..117..393B} to measure the integrated magnitudes of our galaxies and their half-light radii. (Magnitudes from the task to determine $q_{\rm obs}$, i.e. IRAF {\em ellipse}, are not employed because they exclude the contribution of bright clumps, which disable their use in flux determinations.)  We argue that the surface brightness selection effect modeled in Figure~\ref{fig-sb} is not significantly affecting us for three reasons: (1)~Each of our observed $SB$ distributions shows a rough upward trend as $q$ increases, whereas the model populations, which are affected by the selection effect, have a downward trend. (2)~Figure~\ref{fig-sb} shows that only a small percentage of the observed galaxies allows for a selection effect against face-on disks. The selection effect predicts the existence of many galaxies below the 50\% completeness line, which are not observed (cf. top and bottom rows in Figure~\ref{fig-sb})  (3)~Nearly all of our galaxies have some degree of clumpiness, so face-on counterparts may in fact appear brighter than when edge-on due to less extinction of the light from the clumps.  This effect may partially account for the upward trend in each of our observed $SB$ distributions even though the dust content tends to be low in XMPs \citep{2016ApJ...819..110S}.  Differences between the central parts of the galaxies, contributing to $SB$, and the outskirts, contributing to the $q$, could also explain the observed increase of surface brightness with increasing $q$. These two possibilities are discussed in more detail in Appendix~\ref{app:selection}.

 \section{Discussion}
 
We have constrained the 3-D shapes of XMPs from their distribution of projected shapes using a hierarchical Bayesian inference model.  Our main findings (Figure~\ref{fig-clouds}) are that (1) XMPs tend to be triaxial ($A>B>C$) with an intermediate axis $\approx\,$0.7 times the longest axis and that (2) the shortest axis of XMPs ranges between $\approx\,$0.4 and $\approx\,$0.15 times the longest axis, with smaller galaxies tending to be relatively thicker. Our sub-sample cuts based on physical size (Figure~\ref{fig-cuts}) are somewhat arbitrary, but small changes would not significantly affect the inferred range for relative thickness.

To put these results into the proper context, we have to be aware that galaxies of nearly all types have a 3-D shape with some elongation \citep[e.g.,][]{1995A&A...298...63B}, defined as having $B<1$.  Spirals are among the least elongated, with $B \approx 0.9$ on average \citep[e.g.,][]{2008MNRAS.388.1321P}. However, we note that the Milky Way has a stellar halo that is much more elongated than this \citep{2018MNRAS.474.2142I}.  More elongated 3-D shapes (compared to spirals) are common in irregular galaxies
\citep[e.g.,][]{1998ApJ...505..199S,2013MNRAS.436L.104R,2006ApJS..162...49H}, 
including ultra diffuse galaxies\footnote{Using our inference model and the $q$ data in \citet{2017MNRAS.468..703R}, we obtain  $\mu_{B}=0.79\pm 0.06$ and $\mu_{C}=0.49\pm 0.03$. This is in apparent tension with the prolate shape inferred by \citet{2017ApJ...838...93B} for the ultra diffuse galaxies in the Coma cluster. However, the distribution of observed $q$ values are quite similar in both works.}, and in high-redshift, star-forming galaxies \citep[e.g.,][]{2005ApJ...631...85E,2006ApJ...652..963R,2012ApJ...745...85L}.  Simulations have found dark matter halos \citep[e.g.,][]{2002ApJ...574L..21B,2015MNRAS.453..721V,2017MNRAS.467.3226V} and their stellar counterparts \citep[e.g.,][]{2015MNRAS.453..408C,2015MNRAS.453..721V}  also to be notably elongated.

Our results for $C$ are similar to the findings of \citet{2013MNRAS.436L.104R}, who observed relative thickness to increase with decreasing luminosity for dwarf irregular galaxies.  Our average $C$ for sub-sample Large (0.16) is consistent with the relative thickness of the disk component of spiral galaxies \citep[e.g.,][]{2002MNRAS.334..646K}, and our average $C$ for sample All (0.28) agrees closely with \citet{2012ApJ...745...85L} for high-redshift, star-forming galaxies.  \citet{1998ApJ...505..199S} found an average $C$ of 0.55 for blue compact dwarfs (BCDs); this is much greater than our average, yet many XMPs are BCDs \citep[e.g.][]{2000A&ARv..10....1K,2011ApJ...743...77M}.   This discrepancy may reflect in part that XMPs are not a single morphological class (which was evident to us during our sample selection and from the change of surface brightness with $q$).  This issue has been mitigated by omitting obvious spirals and other poorly resolved XMPs and by forming sub-samples based on physical size, but a natural extension of this work would be to evaluate the different morphological types among XMPs.


The trend that smaller galaxies are relatively thicker is understood dynamically since thickness is proportional to the square of the velocity dispersion divided by the disk surface density, which tends to be low for dwarf galaxies like XMPs. If our galaxies all have a similar velocity dispersion and if the smaller ones rotate more slowly, then we would expect the smaller XMPs to be relatively thicker. The ratio of the clump size to the galaxy size also scales with relative thickness because the clump size is usually comparable to the turbulent Jeans length, which scales with velocity dispersion and column density in the same way as the thickness. These scalings make smaller galaxies relatively thicker and more clumpy, which is a morphology that also applies, for the same reasons, to more massive galaxies at high redshift \citep[for further details, see][]{2009ApJ...701..306E}. This tendency for more dynamically primitive systems to be thicker is also found among the faint galaxies in the Virgo and Fornax clusters \citep{2019MNRAS.486L...1S}.  

In addition to slow rotation, large relative thickness may result from feedback ejecting gas and causing the stellar body to expand in response or from stellar scattering, which could be more important in a low mass galaxy with more clumps.  It may also result from the accretion of external gas that induces random motions in the gas that forms stars, which seems to be the case for primitive, high-redshift galaxies \citep[e.g.,][]{2005ApJ...631...85E,2012ApJ...750...95E,2017ApJ...834..181O}.

Isolated galaxies dominated by their own rotation evolve to become axisymmetric disks.  Thus the elongated 3-D shapes of XMPs have to be due to a variety of factors that make them deviate from the ideal case.  According to simulations, larger dark matter halos tend to be more elongated in part because they formed more recently \citep{2017MNRAS.467.3226V}.  Thus, an elongated shape appears to be normal for dynamically young systems in which rotation is probably less dominant.  Lending observational support to this, \citet{2014MNRAS.445.1694L} found that starburst dwarfs with a younger starburst tend to have a more asymmetric H\,I morphology compared to those with an older burst. In this sense, XMPs often have large H\,I reservoirs with asymmetric morphology \citep{2013A&A...558A..18F}. Simulations have also revealed elongated stellar distributions within elongated dark matter halos \citep{2015MNRAS.453..721V}, so that the shape of the stellar distribution is inherited in part from the dark matter halo shape \citep[e.g.,][]{2013ApJ...772..135Z,2014MNRAS.444.3015L}.  The presence of filaments \citep{2017MNRAS.467.3226V}, cosmic accretion \citep{2006ApJ...652..963R}, gas clumps, and radiative feedback \citep{2014MNRAS.442.1545C} may also all contribute to an elongated shape, and such factors are all expected for typical XMPs.  Tidal forces could also cause an elongated shape, but during our sample selection we found a tiny fraction of interacting galaxies with a distorted shape.

The average $SB$ of the XMPs measured in the \rband\  is found to increase as $q$ increases (Figure~\ref{fig-sb}, top row). This is unexpected for disk-like galaxies.  Unfortunately, we do not have a good explanation for this behavior. The ellipse fit employed to infer shapes traces the outer parts of the galaxies, whereas the contribution to the surface brightness is more centrally concentrated. Thus the clumps in XMPs, which we have avoided in the measure of $q$, may bias the trend observed in the $SB$. The unexpected behavior probably indicates that XMPs are not homologous, being more prolate in the centers and more oblate in the outskirts.  This non-homology plus the small contribution of extinction might eventually explain the observed trend.

\acknowledgments
The work has been partly funded by the Spanish Ministry of Economy and Competitiveness (MINECO), projects AYA2016-79724-C4-2-P  (ESTALLIDOS) and AYA2014-60476-P and by the La Caixa Foundation. 
We thank Nicola Caon, Benjamin Alan Weaver, Ignacio Trujillo, Rub\'en S\'anchez-Janssen, and Benne Holwerda for assisting with different aspects of the paper.
We also thank an anonymous referee for suggestions to analyze and clear out the surface brightness bias.
Funding for SDSS-III has been provided by the Alfred P. Sloan Foundation, the Participating Institutions, the National Science Foundation, and the U.S. Department of Energy Office of Science. The SDSS-III web site is http://www.sdss3.org/. SDSS-III is managed by the Astrophysical Research Consortium for the Participating Institutions of the SDSS-III Collaboration including the University of Arizona, the Brazilian Participation Group, Brookhaven National Laboratory, Carnegie Mellon University, University of Florida, the French Participation Group, the German Participation Group, Harvard University, the Instituto de Astrofisica de Canarias, the Michigan State/Notre Dame/JINA Participation Group, Johns Hopkins University, Lawrence Berkeley National Laboratory, Max Planck Institute for Astrophysics, Max Planck Institute for Extraterrestrial Physics, New Mexico State University, New York University, Ohio State University, Pennsylvania State University, University of Portsmouth, Princeton University, the Spanish Participation Group, University of Tokyo, University of Utah, Vanderbilt University, University of Virginia, University of Washington, and Yale University.

\appendix

\section{Demo Axial Ratio Histograms}

\begin{figure}
\begin{center}
\includegraphics[scale = 0.5]{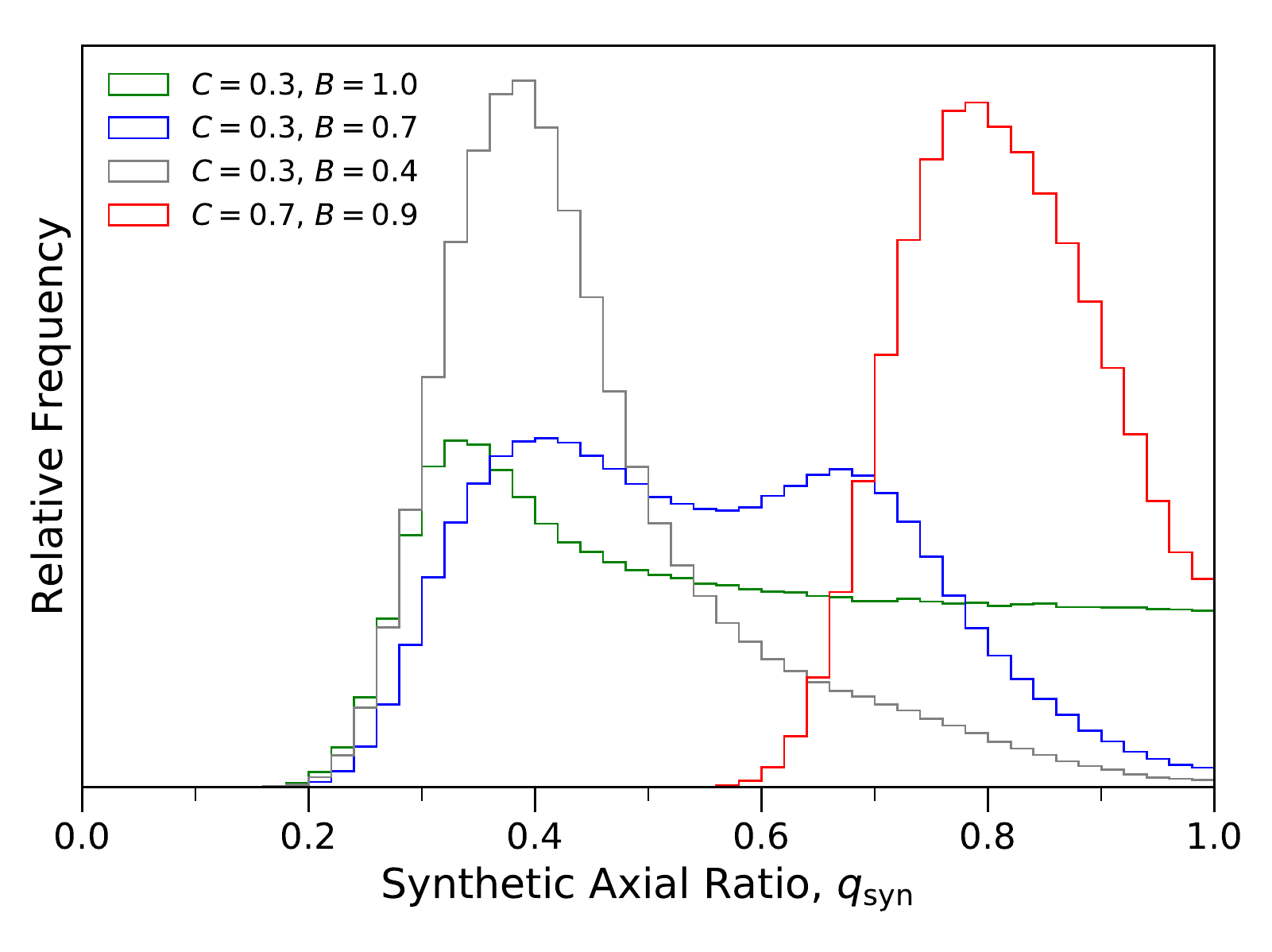} 
\end{center}
\caption{Simulated $q_{{\rm syn}}$ distributions for different 3-D shapes (see Section~2.3 for how we model 3-D shapes). Each galaxy of each model population has the same shape but random orientation. The parameters defining each shape are given in the legend.  Noise was added to $q_{{\rm syn}}$ as explained in Appendix~A.  Each model population contains one million galaxies, and the histogram bin size is 0.02. \label{fig-demo}}
\end{figure}

To highlight the link between $q$ and our model for 3-D shape (see Section~2.3), Figure~\ref{fig-demo} shows a $q_{{\rm syn}}$ histogram for a perfect disk (in green), for a highly triaxial shape (in blue), for a highly elongated shape (in gray), and for a highly spheroidal shape (in red).  Each histogram is from a model population of one million randomly oriented galaxies all having the same shape, defined in the legend of Figure~\ref{fig-demo}.  Gaussian noise (with the standard deviation set to 4\% of $a$) was added to $a$ and $b$, which were then reordered to ensure $q_{{\rm syn}} \leq 1$.

\section{Can the Family Parameters be inferred?}

\begin{figure}
\begin{center}
\includegraphics[scale = 0.5]{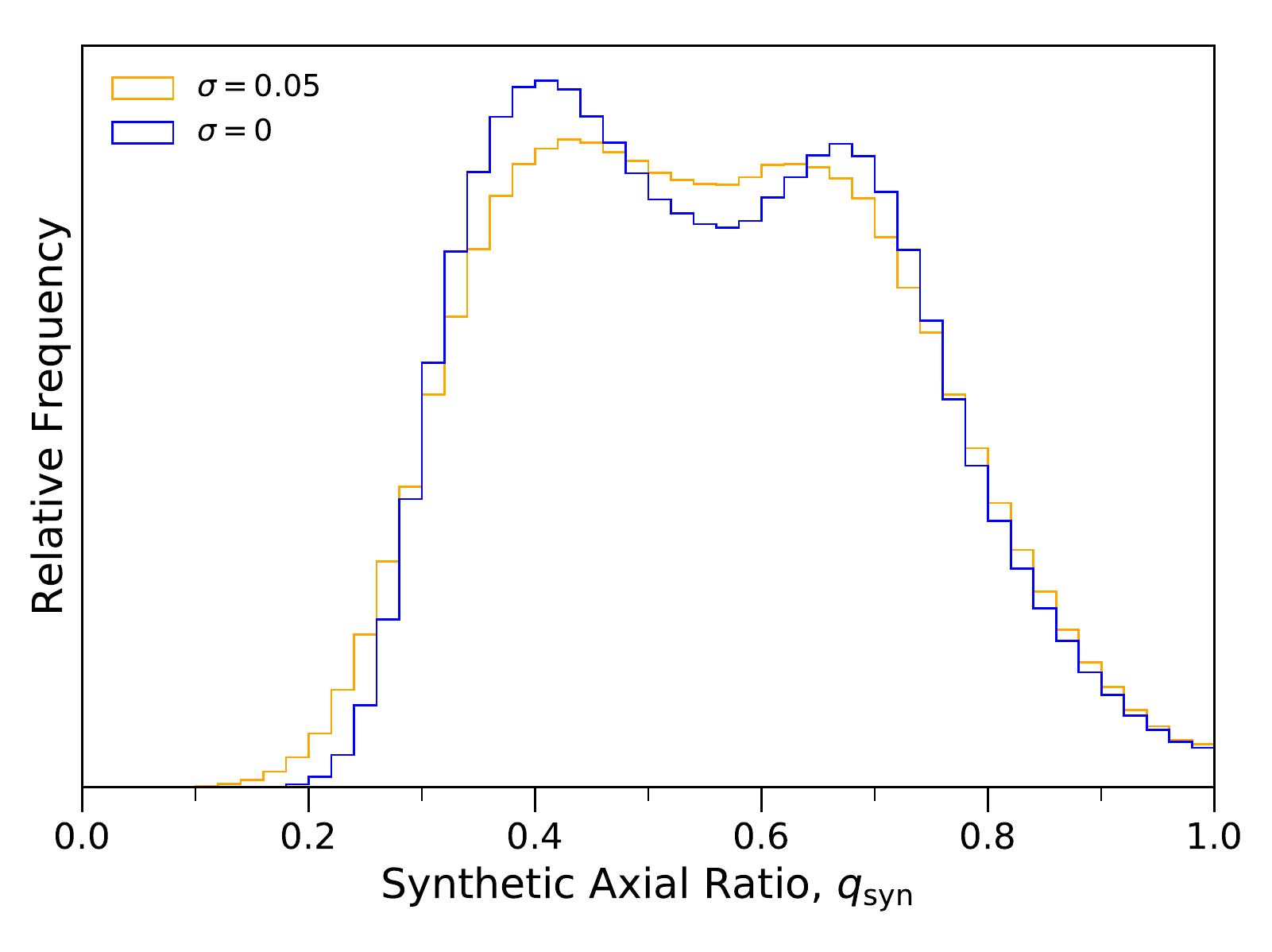} 
\end{center}
\caption{The same blue histogram in Figure~\ref{fig-demo}, now compared to the population when $\sigma_{B}$ and $\sigma_{C}$ (see Section~2.3 for the definition of these parameters) are set to 0.05. \label{fig-sigma}}
\end{figure}

We experimented with $q_{{\rm syn}}$ data of a model population to see how well our inference model can retrieve the set of family parameters \{$\mu_{B}$, $\sigma_{B}$, $\mu_{C}$, $\sigma_{C}$\}.  We tested 10 sub-samples all randomly extracted from the sample of one million galaxies that make up the blue histogram in Figure~\ref{fig-demo}.  Five of the samples have a sample size of 25, the other five samples have a sample size of 200, and all samples have $C=0.3$, $B=0.7$, and noise (added as described in Appendix~A).  If the family parameters are correctly inferred, we should find that the maximum A Posterior solution is consistent with $\mu_C=0.3$, $\sigma_{C}=0$, $\mu_B=0.7$, and $\sigma_{B}=0$.  The inferred family parameter results are summarized in Table~\ref{tbl-tests}.  We can conclude that $\mu_{B}$ and $\mu_{C}$ can be accurately inferred from small samples even though $\sigma_{B}$ and $\sigma_{C}$ are difficult to constrain, and inferring $\sigma_{B}$ remains particularly difficult even as sample size increases.  Another way to see this is in Figure~\ref{fig-sigma}, where the blue histogram in Figure~\ref{fig-demo} is shown again along with the population when $\sigma_{{B}}$ and $\sigma_{{C}}$ are set to 0.05 and with noise in $q_{{\rm syn}}$ (as described in Appendix~A).  Comparing the two histograms, we see that they are virtually identical in the falloff toward higher $q_{{\rm syn}}$, which is where most of the information on $B$ is contained (see Figure~7).  Thus, it makes sense that $\sigma_{{B}}$ remains inherently difficult to constrain even as sample size increases.
\begin{deluxetable}{lccccc}
\tablecolumns{8}
\tablewidth{0pc}
\tablecaption{Maximum A Posteriori Family Parameters: Model Population Tests\tablenotemark{a}\label{tbl-tests}}
\tablehead{
		\colhead{Test No.}   & 
		\colhead{Sample Size}   & 
		\colhead{$\mu_{\rm C}$}    &
		\colhead{$\sigma_{\rm C}$}    &
		\colhead{$\mu_{\rm B}$}    &
		\colhead{$\sigma_{\rm B}$} }  
\startdata
1 & 25 & $0.28^{+0.03}_{-0.04}$ & $0.03^{+0.03}_{-0.02}$ & $0.68^{+0.06}_{-0.05}$ & $0.04^{+0.03}_{-0.02}$ \\
2 & 25 & $0.28^{+0.03}_{-0.03}$ & $0.02^{+0.02}_{-0.01}$ & $0.71^{+0.12}_{-0.21}$ & $0.03^{+0.03}_{-0.02}$ \\
3 & 25 & $0.27^{+0.03}_{-0.03}$ & $0.02^{+0.02}_{-0.01}$ & $0.76^{+0.06}_{-0.06}$ & $0.03^{+0.03}_{-0.02}$ \\
4 & 25 & $0.30^{+0.03}_{-0.03}$ & $0.03^{+0.02}_{-0.02}$ & $0.53^{+0.13}_{-0.07}$ & $0.04^{+0.03}_{-0.02}$ \\
5 & 25 & $0.33^{+0.03}_{-0.04}$ & $0.03^{+0.02}_{-0.02}$ & $0.77^{+0.08}_{-0.10}$ & $0.04^{+0.03}_{-0.02}$ \\
6 & 200 & $0.32^{+0.01}_{-0.01}$ & $0.02^{+0.02}_{-0.01}$ & $0.71^{+0.02}_{-0.02}$ & $0.03^{+0.02}_{-0.01}$ \\
7 & 200 & $0.31^{+0.01}_{-0.01}$ & $0.01^{+0.02}_{-0.01}$ & $0.73^{+0.03}_{-0.03}$ & $0.04^{+0.03}_{-0.02}$ \\
8 & 200 & $0.30^{+0.01}_{-0.01}$ & $0.02^{+0.02}_{-0.01}$ & $0.69^{+0.03}_{-0.03}$ & $0.05^{+0.03}_{-0.03}$ \\
9 & 200 & $0.29^{+0.01}_{-0.01}$ & $0.01^{+0.01}_{-0.01}$ & $0.74^{+0.04}_{-0.03}$ & $0.04^{+0.03}_{-0.02}$ \\
10 & 200 & $0.30^{+0.01}_{-0.01}$ & $0.02^{+0.01}_{-0.01}$ & $0.67^{+0.03}_{-0.03}$ & $0.06^{+0.03}_{-0.03}$ \\
11 & 1000 & $0.30^{+0.00}_{-0.00}$ & $0.01^{+0.01}_{-0.00}$ & $0.98^{+0.01}_{-0.01}$ & $0.00^{+0.00}_{-0.00}$ \\
\enddata

\tablenotetext{a} {We report the median values of the marginal posteriors and express the uncertainty using the 16th and 84th percentiles (the probability that the parameter is within this range is 68\%). The true values are $\mu_C=0.3$, $\sigma_C=0$, $\mu_B=0.7$, and $\sigma_B=0$ for test number 1 to 10. Trial 11 has been run with all other parameters the same but $\mu_B=1.0$.}

\end{deluxetable}

\section{Surface brightness versus axial ratio}\label{app:selection}
Our simple 3-D model is used in Sect.~\ref{sec:selection} to argue that the bias against face-on disk-like galaxies is not present in the analyzed XMP dataset. Rather, galaxies become slightly brighter as $q$ increases (Fig.~\ref{fig-sb}). The 3-D model used to represent galaxy shapes  is simply too elementary to account for the observed effect, however, small modifications of the original model allows us to reproduce the observed trend. The issue is discussed here for consistency, to show that there is no fundamental limitation that prevents our 3-D model from reproducing the trend in Fig.~\ref{fig-sb}.

\begin{figure}
\begin{center}
\includegraphics[scale = 0.9]{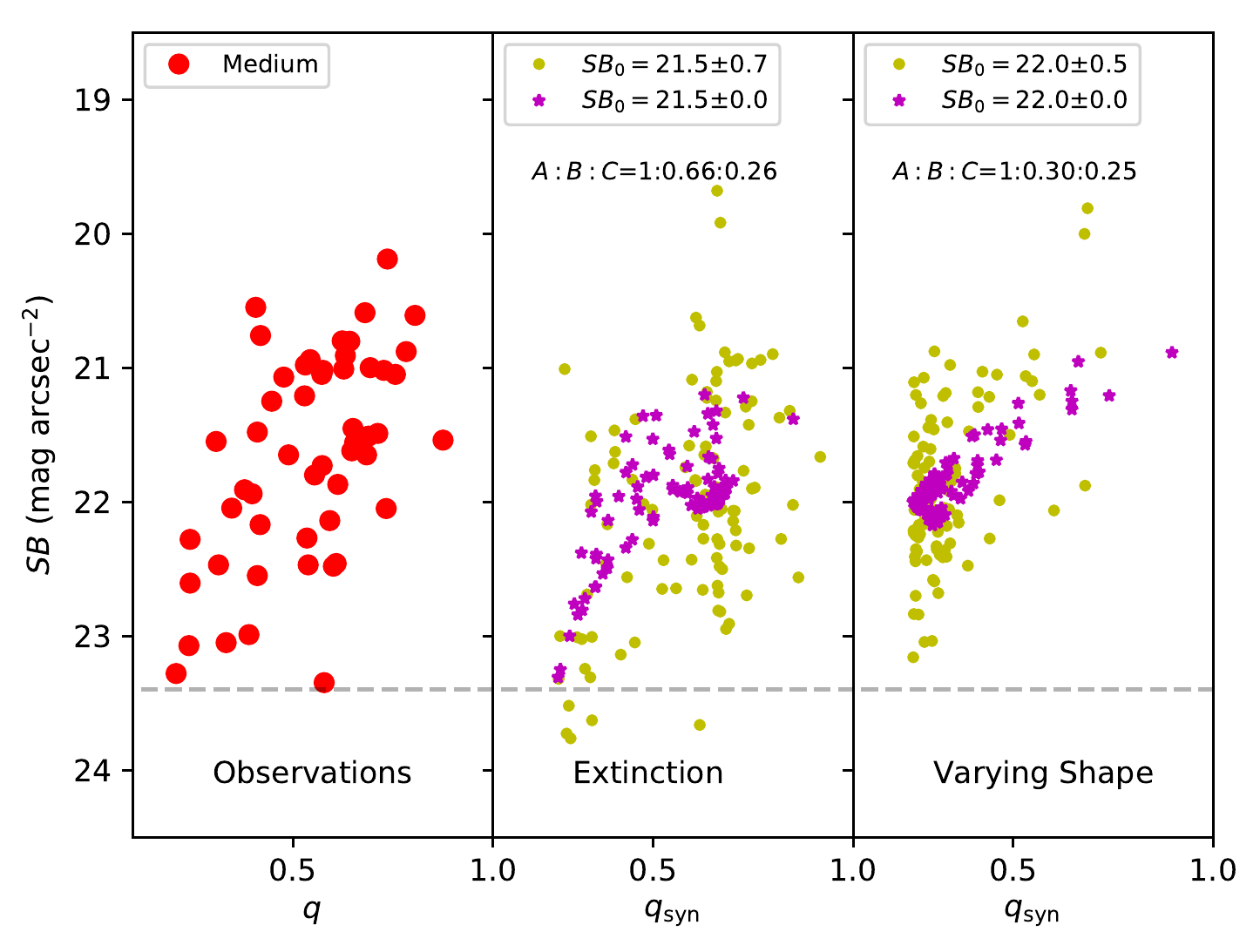} 
\end{center}
\caption{
Left panel: half-light surface brightness in the \rband\ versus $q$ for our sub-sample Medium. The dashed lines show the 50\,\%  completeness level for the SDSS spectroscopic catalog (Sect.~\ref{sec:selection}).  
Center panel: model of the surface-brightness change assuming extinction varying with the galaxy orientation. As observed, model galaxies become brighter with increasing $q_{\rm syn}$ (i.e., SB decreases with increasing $q_{\rm syn}$).
Right panel: model for the surface brightness bias assuming the galaxies to be prolate ellipsoids. Each point corresponds to a galaxy oriented randomly,  with the axial ratio $A:B:C$ and the lowest surface brightness $SB_0$ given in the insets. $SB_0$ is constant for the magenta symbols and has the random spread given in the insets for the yellow symbols. 
\label{fig:appendixonsb}}
\end{figure}

If dust extinction is included in the model, objects observed edge-on may be more obscured and so should present lower surface brightness. The effect of extinction on SB can be easily incorporated into Equation~(\ref{eq:myeq}) adding an extra term, namely,
\begin{equation}
SB = SB_0 -2.5\,\log\big[B/(ab)\big]+2.5\, \kappa\, (q_{\rm syn}^{-1}-1),
\label{eq:myeq_ext}
\end{equation}  
where  $\kappa$ stands for the extinction coefficient, and the dependence $(q_{\rm syn}^{-1}-1)$ is an ansatz to provide a mathematical representation of the attenuation. It was chosen because $(q_{\rm syn}^{-1}-1)$ is zero for face-on disks ($q_{\rm syn}=1$) and increases as $q^{-1}$ with decreasing $q_{\rm syn}$, which is the behavior expected for thin disks, for which $q_{\rm syn}$ is just the cosine of the inclination with respect to the line-of-sight. Figure~(\ref{fig:appendixonsb}), center panel, shows the same type of Monte Carlo simulation worked out in Sect.~\ref{sec:selection} but now based on Equation~(\ref{eq:myeq_ext}). We begin with 100 randomly oriented galaxies, for which $ab$ and $q_{\rm syn}$ are computed from Equations (\ref{eq:area}) to (\ref{eq:qsyn}). These galaxies are then selected or not with a probability set by completeness function (see Sect. ~\ref{sec:selection}). We also make two assumptions regarding $SB_0$, i.e., it is the same for all galaxies (the magenta symbols in Fig.~\ref{fig:appendixonsb}), or they have a random spread of values of $\pm 0.7$ mag (the yellow symbols in Fig.~\ref{fig:appendixonsb}).  We use $\kappa =0.3$, which is a reasonable value for XMPs 
\citep[see, Sect.~4.1 in][where $\kappa$ has been ascribed to the extinction coefficient in H$\beta$]{2016ApJ...819..110S}.  As Fig.~\ref{fig:appendixonsb} evidences,  including extinction produces an increase SB as $q_{\rm syn}$ decreases, in qualitative agreement with the observed trend (cf. Figs.~\ref{fig:appendixonsb}, left and center panels).

Another alternative explanation relies on the inside-out change in galaxy shape. The increase of surface brightness with increasing $q_{\rm syn}$ is to be expected in 3-D {\em prolate} ellipsoids, which have $A > B, C$ and $B\approx C$. According to Equation~(\ref{eq:myeq}), the galaxies are brightest when their projected area is smallest. In the case of a prolate ellipsoid, the minimum area on the sky results when the galaxy is observed along its major axis, and since $B\approx C$, this maximum brightness corresponds to $q\simeq 1$. The behavior is shown in  Figure~\ref{fig:appendixonsb}, right panel, which follows from ellipsoids having $A:B:C = 1:0.30:0.25$. This prolate shape disagrees with the oblate shape inferred from the $q$~technique. Thus, for this to be an explanation,  the 3-D shape of the galaxies should be non-homologous, being more prolate in the centers and more oblate in the outskirts. The ellipse fit used to infer shapes traces the outer parts of the galaxies, whereas the contribution to the surface brightness is more centrally concentrated. This non-homology might explain the trend observed in $SB(q)$. The points in Figure~\ref{fig:appendixonsb}, right panel, make two assumptions regarding $SB_0$: it is the same for all galaxies (the magenta symbols), or they have a random spread of values of $\pm 0.5$ mag (the yellow symbols). The actual values of $SB_0$ have been tuned to match the level and  spread of the observed SB (Figure~\ref{fig:appendixonsb}, left panel).



\end{document}